\documentclass[
prd
,byrevtex
,preprint
,amsfonts,amsmath,amssymb
]{revtex4}
\usepackage{bm}
\usepackage[dvipdfmx]{graphicx}
\usepackage{ascmac}
\usepackage{latexsym}
\usepackage{physics}
\usepackage{color,float}
\begin{document}
\title{Field-induced entanglement in spatially superposed objects}
\author{Akira Matsumura}
\email{matsumura.akira@phys.kyushu-u.ac.jp}
\affiliation{Department of Physics, Kyushu University, Fukuoka, 819-0395, Japan}

\begin{abstract}
We discuss the generation of field-induced entanglement between two objects each in a superposition of two trajectories. 
The objects have currents coupled to local quantum fields, and the currents are evaluated around each trajectory of the objects.  
The fields have only dynamical degrees of freedom and satisfy the microcausality condition. 
We find that the superposed state of trajectories cannot be entangled when the objects are spacelike separated. 
This means that the quantum fields do not generate spacelike entanglement in the superposition of two trajectories of each object. 
\end{abstract}
\maketitle

\tableofcontents
\section{Introduction}

The full picture of quantum gravity 
\cite{Feynmann1995, Aharony2000, Kiefer2006, Woodard2009}, 
which unifies general relativity and quantum mechanics, is still unclear.
This is attributed to the lack of theoretical and experimental approaches to connect gravitational and quantum phenomena.
However, with the recent development of various quantum technologies 
\cite{Doherty2013, Aspelmeyer2014, Matsumoto2019, Catano-Lopez2020}, 
there have been attempts to clarify quantum natures of gravity (for example, see 
\cite{Carney2018} 
and the references therein, or the recent works 
\cite{Marletto2018, Carlesso2019, Krisnanda2020, vandeKamp2020, Grobardt2020, Qvafort2020, Guerreiro2020, Kanno2020, Hall2018, Belenchia2018, Howl2019, Marshman2020, Anastopoulos2020, Balushi2018, Miao2020, Matsumura2020, Nguyen2020, Miki2020, Tilly2021}).
In such works,
the quantum gravity induced entanglement of masses (QGEM) proposal
\cite{Bose2017, Marletto2017, Marshman2020}
has been attracting attentions.
In the proposal, the authors considered two objects each in a superposition of two trajectories and assumed the Newtonian potential between them. 
The gravitational interactions generate the entanglement between the two objects. 
The detection of gravity-induced entanglement can be a witness of quantum nature of gravity. 

The interesting point in the QGEM proposal is that two spatially superposed objects can probe quantum entanglement induced by fields. 
This is analogous to entanglement harvesting protocols 
\cite{Reznik2003, Lin2010, Salton2015, Pozas-Kerstjens2015, Pozas-Kerstjens2016, Simidzija2018a, Stritzelberger2021, Henderson2020, Foo2021} 
by the Unruh-DeWitt detector.
The Unruh-DeWitt detector is constructed by a particle with internal degrees of freedom, which locally interacts with a quantum field. 
In this model, the source of entanglement is the quantum field. 
In particular, it is known that the spacelike entanglement of a vacuum state induces the entanglement between the two spatially separated detectors (for example, see \cite{Reznik2003}).

In this paper, 
we investigate how two superposed objects are capable to probe the entanglement of quantum fields.
We assume that the fields have only dynamical degrees of freedom and any constraint equations are not imposed on the whole Hibert space of the objects and the fields.
We consider the superposed objects which do not interact with each other and whose currents locally couple with the fields.
By evaluating the currents along the objects' trajectory, we compute the time evolution of the total system.  
For the case where the objects are spatially separated, we show that the state of trajectories remain disentangled if the microcausality condition holds for the quantum fields. 
In other words, such quantum fields cannot be mediators of spacelike entanglement for superposed trajectories of the objects.
Our analysis also presents possible approaches and extensions of the objects' model to verify the spacelike entanglement of fields; use of the internal degrees of freedom and extended model with multi-objects or multi-trajectories.

This paper is organized as follows. 
In Sec. \ref{sec:BMVproposal}, the QGEM proposal to test quantum gravity and its thoretical approach are reviewed. 
In Sec. \ref{sec:ourmodel}, we introduce the model with the interaction given in a bilinear form of fields and currents of two objects. 
We derive the solution of the Schr$\ddot{\text{o}}$dinger equation. 
In Sec. \ref{sec:Noentangle}, we investigate the separability of the two objects based on the solution.  
We find the no-go result of generation of spacelike entanglement, and discuss its implications.  
In Sec. \ref{sec:Conclusion}, the conclusion is devoted. 
We use the natural units 
$\hbar=c=1$ in this paper. 


\section{Quantum gravity induced entanglement of masses
}
\label{sec:BMVproposal}

\begin{figure}[htbp]
  \centering
  \includegraphics[width=0.50\linewidth]{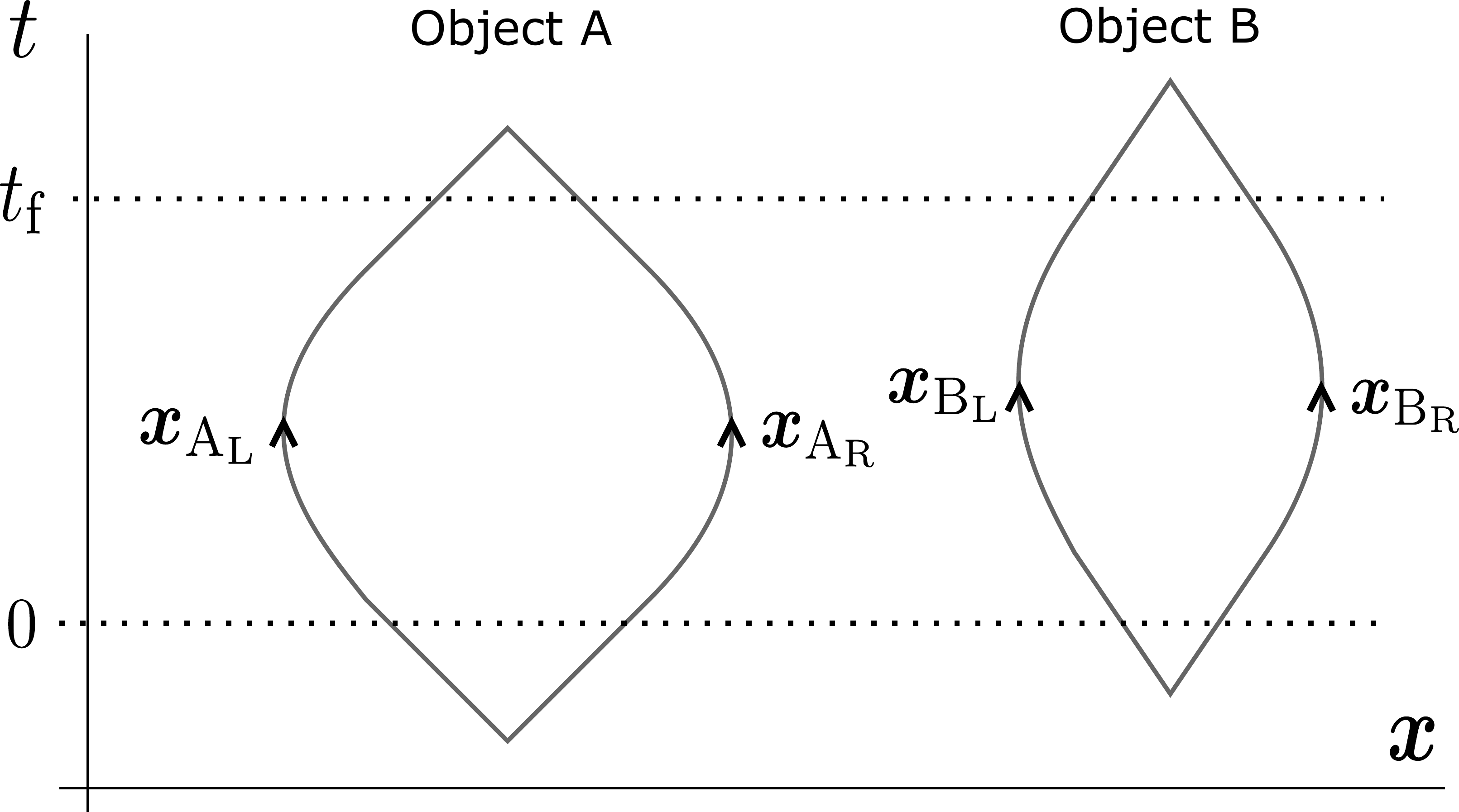}
  \caption{A configuration of the trajectories of the objects A and B. For the QGEM proposal, the entanglement is generated between the objects by the gravitational interaction.}
  \label{fig:trajectory}
\end{figure}

The experimental setting of two matter-wave interferometers to test quantum gravity was proposed, which is called the QGEM proposal \cite{Bose2017, Marletto2017, Marshman2020}. 
In each interferometer, a single object is in a superposition of two trajectories. 
Fig.\ref{fig:trajectory} presents a rough configuration of trajectories of each object. 
We assume that the two objects interact with each other by the Newtonian potential. 
The Hamiltonian of the objects is 
\begin{align}
\hat{H}_\text{QGEM}
=\hat{H}_\text{A}+\hat{H}_\text{B}+\hat{V}_\text{AB}, 
\quad 
\hat{V}_\text{AB}= -\frac{Gm_\text{A} m_\text{B}}{|\hat{\bm{x}}_\text{A} - \hat{\bm{x}}_\text{B}|},
\label{eq:HBMV}
\end{align}
where 
$m_\text{A}$ and 
$m_\text{B}$ are the masses of the objects A and B, 
$\hat{\bm{x}}_\text{A}$ and 
$\hat{\bm{x}}_\text{B}$ are each position, and the Hamiltonian 
$\hat{H}_\text{A}$ and 
$\hat{H}_\text{B}$ determine each trajectory of the objects. 
Each of the two objects at 
$t=0$ is in the spatially superposed state, 
\begin{equation}
| \psi_\text{in} \rangle 
= 
\frac{1}{\sqrt{2}} (|\psi_\text{R} \rangle_\text{A} + |\psi_\text{L} \rangle_\text{A}) 
\otimes 
\frac{1}{\sqrt{2}} (|\psi_\text{R}\rangle_\text{B} + |\psi_\text{L} \rangle_\text{B}),
\label{eq:psiinBMV}
\end{equation}
where 
$|\psi_\text{R} \rangle_\text{A}$
and 
$|\psi_\text{L} \rangle_\text{A}$ 
are the states with wave packets localized around positions 
$\bm{x}=\bm{x}_{\text{A}_\text{R}}(0)$ 
and 
$\bm{x}=\bm{x}_{\text{A}_\text{L}}(0)$ at 
$t=0$, respectively.
Also, 
$|\psi_\text{R} \rangle_\text{B}$
and 
$|\psi_\text{L} \rangle_\text{B}$ are defined in the same manner.
Those states 
satisfy
${}_\text{A} \langle \psi_\text{R} | \psi_\text{L} \rangle_\text{A} \approx0$ and  
${}_\text{B} \langle \psi_\text{R} | \psi_\text{L} \rangle_\text{B} \approx0$ when each wave packet is sufficiently separated.
The evolved state 
$|\psi_\text{f} \rangle$ at 
$t=t_\text{f}$ is 
\begin{align}
| \psi_\text{f}  \rangle 
&
= e^{-i t_\text{f} \hat{H}_\text{QGEM} } | \psi_\text{in} \rangle
\nonumber 
\\
&
= e^{-i t_\text{f} (\hat{H}_\text{A} + \hat{H}_\text{B})  } \text{T} \exp \Bigl[ i \int^{t_\text{f}}_0 dt \frac{Gm_\text{A} m_\text{B}}{|\hat{\bm{x}}^\text{I}_\text{A} (t)-\hat{\bm{x}}^\text{I}_\text{B} (t) | }\Bigr] | \psi_\text{in} \rangle 
\nonumber 
\\
&
\approx \frac{1}{2} e^{-i t_\text{f} (\hat{H}_\text{A} + \hat{H}_\text{B})  } 
\sum_{\text{P}, \text{Q}=\text{R},\text{L}} e^{i \Phi_\text{PQ}} |\psi_\text{P} \rangle_\text{A}  |\psi_\text{Q} \rangle_\text{B}, 
\label{eq:psiBMV}
\end{align}
where $\text{T}$ is the time-ordered product, and 
$\hat{\bm{x}}^\text{I}_\text{A} (t)=e^{it(\hat{H}_\text{A}+\hat{H}_\text{B})} \hat{\bm{x}}_\text{A} e^{-it(\hat{H}_\text{A}+\hat{H}_\text{B})}$ and
$\hat{\bm{x}}^\text{I}_\text{B} (t)=e^{it(\hat{H}_\text{A}+\hat{H}_\text{B})} \hat{\bm{x}}_\text{B} e^{-it(\hat{H}_\text{A}+\hat{H}_\text{B})}$ are the position operators in the interaction picture.
The phase shift
\begin{equation}
\Phi_\text{PQ} 
= \int^{t_\text{f}}_0 dt \frac{Gm_\text{A} m_\text{B}}{|\bm{x}_{\text{A}_\text{P}}(t)-\bm{x}_{\text{B}_\text{Q}} (t)| }
\label{eq:PhiPQ}
\end{equation}
is given by the Newtonian potential between the two objects on the trajectories 
$\bm{x}=\bm{x}_{\text{A}_\text{P}}(t)$ and 
$\bm{x}=\bm{x}_{\text{B}_\text{Q}}(t)$ 
$(\text{P}, \text{Q} =\text{R},\text{L})$. 
In the expression \eqref{eq:psiBMV}, we omitted the symbol of the tensor product as
$|\, \cdot \, \rangle_\text{A} \otimes |\, \cdot \, \rangle_\text{B} 
=|\, \cdot \, \rangle_\text{A}  |\, \cdot \, \rangle_\text{B}$.
The approximation of the third line of Eq. \eqref{eq:psiBMV} is given as 
\begin{equation}
\hat{\bm{x}}^\text{I}_\text{A} (t) | \psi_\text{P} \rangle_\text{A} 
\approx \bm{x}_{\text{A}_\text{P}}(t) |\psi_\text{P} \rangle_\text{A}, 
\quad
\hat{\bm{x}}^\text{I}_\text{B} (t) |\psi_\text{Q} \rangle_\text{B} 
\approx \bm{x}_{\text{B}_\text{Q}}(t) |\psi_\text{Q} \rangle_\text{B}. 
\label{eq:approx}
\end{equation}
These equations are valid when the size of each wave packet is larger than de Broglie wave length of each object \cite{Ford1993, Breuer2001}.
Choosing the masses, the distance between a pair of trajectories and the traveling time properly, we find that the state \eqref{eq:psiBMV} is entangled. 
Hence the gravitational interaction can generate quantum entanglement. 
The key point in the QGEM proposal is that the spatially superposed objects can probe quantum entanglement. 
In the following sections, we will discuss the detection of entanglement of dynamical fields by using such objects. 

\section{Model Hamiltonian for fields and objects}
\label{sec:ourmodel}

In this section, we introduce a model of two obhects and fields to examine the detection of entanglement of the fields.
In the Schr$\ddot{\text{o}}$dinger picture, we consider the Hamiltonian of two objects A and B and fields as 
\begin{align}
\hat{H}
=\hat{H}_\text{A}+\hat{H}_\text{B}+\hat{H}_\text{F}+\hat{V}, 
\quad 
\hat{V}= \int d^3 x \,(\hat{\bm{J}}_\text{A} (\bm{x})+\hat{\bm{J}}_\text{B} (\bm{x})) \cdot \hat{\bm{\phi}} (\bm{x}), 
\label{eq:Htot}
\end{align}
where the Hamiltonians
$\hat{H}_\text{A}$, 
$\hat{H}_\text{B}$ and 
$\hat{H}_\text{F}$ determine each dynamics of the objects A, B and the fields. 
The vectors $\hat{\bm{J}}_\text{A}$ 
and 
$\hat{\bm{J}}_\text{B}$ are current operators with respect to the objects A and B, and 
$\hat{\bm{\phi}}$ is the field operator. 
The inner product 
$\bm{J} \cdot \bm{\phi}$ is 
defined by 
$\sum_{k} J^k \phi_k$ with labels 
$k$. 

We assume that the fields have only dynamical degrees of freedom and there are no constraint equations on the whole Hilbert space. 
The field operators are represented on a physical Hilbert space 
$\mathcal{H}_\text{F}$ without negative norm states.  
In gauge field theories, there are formalisms using an unphysical Hilbert space of fields with gauge degrees of freedom \cite{Weinberg1996}. 
The fact that there are no negative norm states will be used to derive our result in the next section.  

We note that the Hamiltonian \eqref{eq:Htot} does not completely represent that in the linearized Einstein theory. 
At first glance, by choosing the component of currents 
$\hat{J}^k_\text{A}$,
$\hat{J}^k_\text{B}$ and
the fields 
$\hat{\phi}_k$ as the energy-momentum tensor 
$\hat{T}^{\mu \nu}$ and 
the metric perturbation 
$\hat{h}_{\mu \nu}$ properly, 
the local interaction 
$\hat{V}$ seems to be that in the linearized Einstein theory. 
This is not correct since the fields and those Hilbert space
$\mathcal{H}_\text{F}$ are assumed not to have gauge degrees of freedom and negative norm states. 
Also, even for the transverse traceless gauge 
($\hat{h}_{\mu \nu}$ have only physical modes), the Hamiltonian 
\eqref{eq:Htot} is not admitted in the linearized Einstein theory. 
This is because, from the constraints of the Einstein equation, the non-dynamical parts of the metric perturbation give nonlocal interactions such as the Newtonian potential. 
However, there are no nonlocal interactions between the two objects in our model.

The almost same argument holds for the quantum electromagnetic dynamics, but we can admit an effective model described by the Hamiltonian \eqref{eq:Htot}. 
Let us consider that the objects A and B without total electric charges and with the electric dipole moments 
$\hat{\bm{d}}_\text{A}$ and 
$\hat{\bm{d}}_\text{B}$, respectively.
For the distant objects, the Coulomb potential between them is neglected, and the local coupling to an electric field 
$\hat{\bm{E}}$ can be dominant. 
By assigning the field operator 
$\hat{\bm{\phi}}$ 
and the currents
$\hat{\bm{J}}_\text{A} $ and 
$\hat{\bm{J}}_\text{B}$ to 
$ \hat{\bm{E}}$,
$\hat{\bm{d}}_\text{A} \delta^3 (\bm{x}-\bm{x}_\text{A})$
and 
$ \hat{\bm{d}}_\text{B} \delta^3 (\bm{x}-\bm{x}_\text{B})$,
our model describes the objects with the dipole coupling to the electric field at the positions
$\bm{x}=\bm{x}_\text{A}$ and 
$\bm{x}=\bm{x}_\text{B}$. 
In \cite{Pozas-Kerstjens2016}, a similar model with time-dependent couplings and spatially smearing functions was considered as the Unruh-DeWitt detector model.

We consider that each object at $t=0$ is 
in a superposition of two local states
$|\psi_\text{R} \rangle $ and 
$|\psi_\text{L} \rangle $ with 
$\langle \psi_\text{P} | \psi_{\text{P}'} \rangle \approx \delta_{\text{PP}'}$ 
($\text{P}, \text{P}'=\text{R}, \text{L}$).
As the mentioned above, the interation of the model \eqref{eq:Htot} can describe dipole coupling in the quantum electrodynamics. 
Each object may have some internal degrees of freedom such as electric dipole moments. 
We assume that the internal degrees of freedom of each objects at 
$t=0$ is in states 
$|a \rangle_\text{Ai}$ and 
$|b \rangle_\text{Bi}$, respectively. 
The objects move on the trajectories determined by the Hamiltonian 
$\hat{H}_\text{A}$ and 
$\hat{H}_\text{B}$ (see Fig. \ref{fig:trajectory}).  
The current operators
$\hat{\bm{J}}^{\text{I}}_\text{A} (t,\bm{x})=e^{i\hat{H}_0 t} \hat{\bm{J}}_\text{A} (\bm{x}) e^{-i\hat{H}_0 t}$ and 
$\hat{\bm{J}}^{\text{I}}_\text{B} (t, \bm{x})=e^{i\hat{H}_0 t} \hat{\bm{J}}_\text{B} (\bm{x}) e^{-i\hat{H}_0 t}$ in the interaction picture defined with $\hat{H}_0 = \hat{H}_\text{A} +\hat{H}_\text{B} +\hat{H}_\text{F}$ are approximated by the local values :
\begin{equation}
\hat{\bm{J}}^{\text{I}}_{\text{A}, \text{I}} (t,\bm{x}) |\psi_\text{P} \rangle_\text{A} \otimes   | a \rangle_\text{Ai} 
\approx  |\psi_\text{P} \rangle_\text{A} \otimes  \, \hat{\bm{j}}^\text{I}_{\text{A}_\text{P}}(t,\bm{x}) | a \rangle_\text{Ai}, \quad 
\hat{\bm{J}}^{\text{I}}_\text{B} (t,\bm{x}) |\psi_\text{Q} \rangle_\text{B} \otimes  | b \rangle_\text{Bi} 
\approx |\psi_\text{Q} \rangle_\text{B} \otimes \, \hat{\bm{j}}^\text{I}_{\text{B}_\text{Q}}(t,\bm{x})  |b \rangle_\text{Bi},
\label{eq:jAjB}
\end{equation}
where  
$\hat{\bm{j}}^\text{I}_{\text{A}_\text{P}}(t,\bm{x})=\hat{\bm{s}}^\text{I}_\text{A} (t) \delta^3 (\bm{x}-\bm{x}_{\text{A}_\text{P}} (t) )$ 
and 
$\hat{\bm{j}}^\text{I}_{\text{B}_\text{Q}}(t,\bm{x})=\hat{\bm{s}}^\text{I}_\text{B} (t) \delta^3 (\bm{x}-\bm{x}_{\text{B}_\text{Q}} (t) )$ 
($\text{P},\text{Q}=\text{R},\text{L}$) 
with the internal physical quantities
$\hat{\bm{s}}^\text{I}_\text{A} (t)$ and 
$\hat{\bm{s}}^\text{I}_\text{B} (t)$ acting on the Hilbert spaces 
$\mathcal{H}_\text{Ai}$ and 
$\mathcal{H}_\text{Bi}$ of internal degrees of freedom, respectively.
For example, if the objects have electric dipole moments and the fields are electric fields, the classical current 
$\hat{\bm{j}}^\text{I}_{\text{A}_\text{P}}(t,\bm{x})$ of the object A has the form 
$\hat{\bm{j}}^\text{I}_{\text{A}_\text{P}}(t,\bm{x}) = \hat{\bm{d}}^{\text{I}}_\text{A}(t) \delta^3 (\bm{x}-\bm{x}_{\text{A}_\text{P}} (t))$ with the electric dipole 
$\hat{\bm{d}}^\text{I}_\text{A} (t) (= \hat{\bm{s}}^\text{I}_\text{A} (t) )$ in the interaction picture. 
The similar argument is made for the object B. 

When the fields are in a state 
$|\chi \rangle_\text{F}$ at 
$t=0$,  
the state of the objects and the fields at 
$t=0$ is 
\begin{equation}
| \Psi_\text{in} \rangle 
= 
|\alpha \rangle_{\text{A} \otimes \text{Ai}} \, 
|\beta \rangle_{\text{B} \otimes \text{Bi}} \, 
 | \chi \rangle_\text{F}, 
\label{eq:Psiin} 
\end{equation}
where 
\begin{equation}
|\alpha \rangle_{\text{A} \otimes \text{Ai}}
= 
(\alpha_\text{R} |\psi_\text{R}\rangle_\text{A} + \alpha_\text{L} |\psi_\text{L} \rangle_\text{A} ) \otimes |a \rangle_\text{Ai},
\quad
|\beta \rangle_{\text{B} \otimes \text{Bi}}
=(\beta_\text{R} |\psi_\text{R}\rangle_\text{B} + \beta_\text{L} |\psi_\text{L} \rangle_\text{B}) \otimes |b \rangle_\text{Bi}
\label{eq:alphabeta}
\end{equation} 
where 
$|\alpha_\text{R}|^2+|\alpha_\text{L}|^2 \approx 1$ and 
$|\beta_\text{R}|^2+|\beta_\text{L}|^2 \approx 1$ holds since 
the state 
$| \psi_\text{P}\rangle$
satisfies 
$\langle \psi_\text{P}| \psi_{\text{P}'} \rangle \approx \delta_{\text{PP}'}$. 
Note that the initial product state may be not valid if there are constraint equations on the objects and fields. 
The solution of the Schr$\ddot{\text{o}}$dinger equation is 
\begin{align}
| \Psi_\text{f} \rangle 
&
= e^{-i \hat{H}t_\text{f}} | \Psi_\text{in} \rangle
\nonumber \\
&
= e^{-i \hat{H}_0 t_\text{f} } \text{T} \exp \Bigl[ -i \int^{t_\text{f}}_0 dt \int d^3 x \, (\hat{\bm{J}}^{\text{I}}_\text{A} (t,\bm{x})+\hat{\bm{J}}^{\text{I}}_\text{B} (t,\bm{x})) \cdot \hat{\bm{\phi}}^\text{I} (t,\bm{x}) \Bigr] | \Psi_\text{in} \rangle
\nonumber \\
&
\approx 
e^{-i \hat{H}_0 t_\text{f} }  
\sum_{\text{P}, \text{Q}=\text{R},\text{L}} 
\alpha_\text{P} \, \beta_\text{Q} |\psi_\text{P} \rangle_\text{A} | \psi_\text{Q} \rangle_\text{B} 
\, \otimes \hat{U}_\text{PQ}| \chi \rangle_\text{F}  | a \rangle_\text{Ai}  | b \rangle_\text{Bi} ,
\label{eq:Psit}
\end{align}
where 
$\hat{\bm{\phi}}^\text{I} (t,\bm{x})=e^{i\hat{H}_0 t} \hat{\bm{\phi}}(\bm{x}) e^{-i\hat{H}_0 t}$. 
In the third line, we used the approximations \eqref{eq:jAjB} assigning the local currents and defined the unitary operator
\begin{equation}
\hat{U}_\text{PQ}=\text{T} \exp \Bigl[ -i \int^{t_\text{f}}_0 dt \int d^3 x \,
(\hat{\bm{j}}^\text{I}_{\text{A}_\text{P}} (t,\bm{x})+ \hat{\bm{j}}^\text{I}_{\text{B}_\text{Q}} (t,\bm{x})) \cdot \hat{\bm{\phi}}^\text{I} (t,\bm{x}) \Bigr] .
\label{eq:U}
\end{equation}
The unitary operator 
$\hat{U}_\text{PQ}$ acts 
on not only fields' state 
$|\chi \rangle_\text{F} $ but also the states of the internal degrees of freedom of the objects 
$|a \rangle_\text{Ai}$ and 
$|b \rangle_\text{Bi}$. 

In the next section, we examine the entanglement between the two objects A and B using the solution Eq. \eqref{eq:Psit}. 
We will show no generation of entanglement for the trajectories of objects which are in spacelike regions. 
This argument follows by the microcausality of fields, which is independent of the dynamics of the fields.


\section{No generation of spacelike entanglement between two objects}
\label{sec:Noentangle}

In this section, we investigate the generation of entanglement between the two objects. 
Before mentioning our result, we focus on two origins of the generation of entanglement. 

\begin{figure}[htbp]
  \centering
  \includegraphics[width=0.70\linewidth]{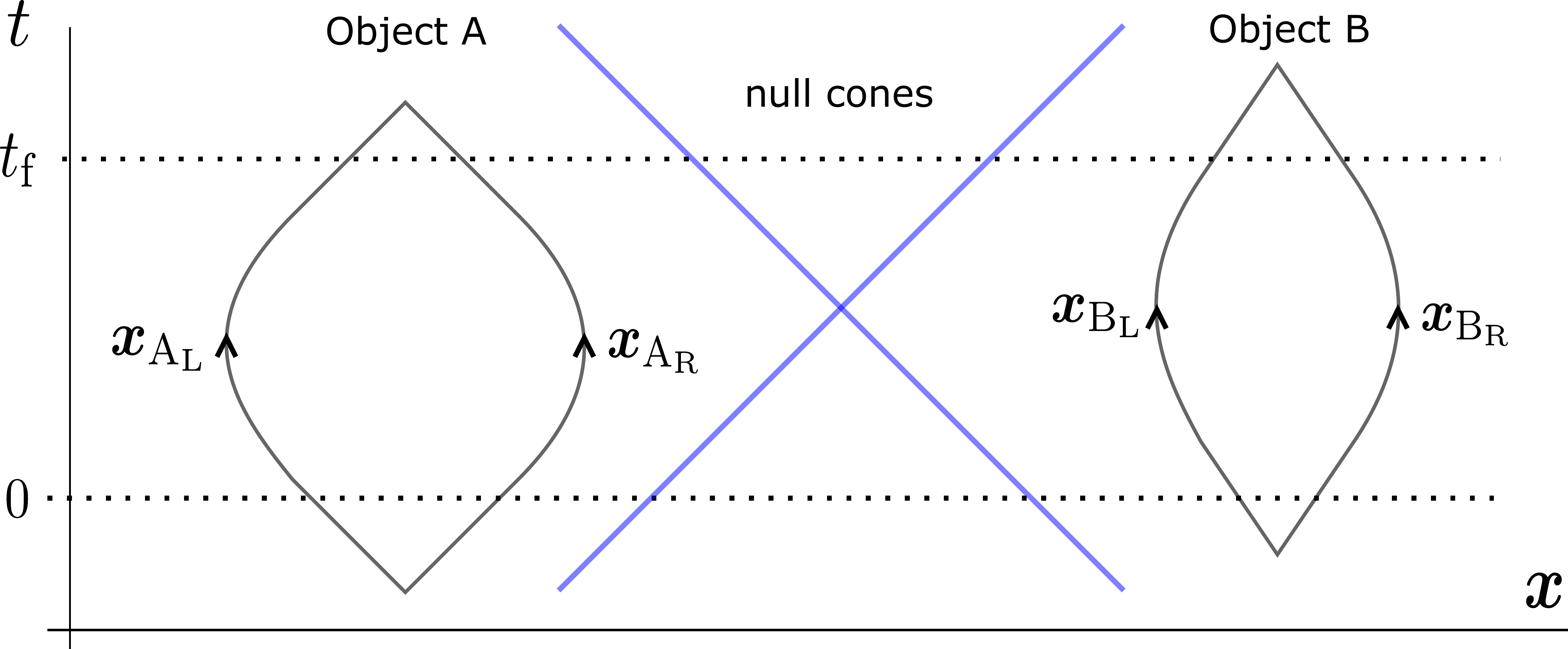}
  \caption{A configuration of trajectories of each object, which is in spatially separated regions. }
  \label{fig:trajectory2}
\end{figure}

First, it is important to consider whether the unitary evolution gives correlations between the objects or not. 
The Hamiltonian 
$\hat{H}_0 =\hat{H}_\text{A} +\hat{H}_\text{B} + \hat{H}_\text{F}$ yields independent dynamics of each system, which give no correlations. 
On the other hand, the unitary evolution 
$\hat{U}_\text{PQ}$ Eq.\eqref{eq:U} given by the local interaction
$\hat{V}$ leads to the following process: 
the object A locally excites the fields, and then the excitaions propagate to the object B and alter the potential around it. 
This process gives effective interactions and induces correlations between the objects A and B. 
In fact, there are no such effects when the two objects are in spacelike separated regions (see Fig. \ref{fig:trajectory2}). 
If the fields in spacelike regions commute each other (the microcausality condition, for example, see \cite{Weinberg1995}), we have 
\begin{equation}
\Bigl[ \int d^3 x \, 
\hat{\bm{j}}^\text{I}_{\text{A}_\text{P}} (t,\bm{x}) \cdot \hat{\bm{\phi}}^\text{I} (t,\bm{x}), \, 
 \int d^3 y \, \hat{\bm{j}}^\text{I}_{\text{B}_\text{Q}} (t',\bm{y}) \cdot \hat{\bm{\phi}}^\text{I} (t',\bm{y}) \Bigr]=0, 
\label{eq:comm}
\end{equation}
where note that 
$\hat{\bm{j}}^\text{I}_{\text{A}_\text{P}} (t,\bm{x}) = \hat{\bm{s}}^\text{I}_\text{A} (t) \delta^3 (\bm{x}-\bm{x}_{\text{A}_\text{P}} (t) ) $ and 
$\hat{\bm{j}}^\text{I}_{\text{B}_\text{Q}} (t',\bm{y}) = \hat{\bm{s}}^\text{I}_\text{B} (t')\delta^3 (\bm{y}-\bm{x}_{\text{B}_\text{Q}} (t'))$ 
with the internal quantities 
$\hat{\bm{s}}^\text{I}_\text{A} (t)$
and 
$\hat{\bm{s}}^\text{I}_\text{B} (t)$ of each object. Then the unitary operator 
$\hat{U}_\text{PQ}$ is factorized into the local unitaries,
\begin{equation}
\hat{U}_\text{PQ}
=
\hat{U}_{\text{A}_\text{P}}  \otimes \hat{U}_{\text{B}_\text{Q}}, 
\label{eq:Udec}
\end{equation}
where
$\hat{U}_{\text{A}_\text{P}}$ and 
$\hat{U}_{\text{B}_\text{Q}}$ are 
\begin{align}
\hat{U}_{\text{A}_\text{P}}
&=\text{T} \exp \Bigl[ -i \int^{t_\text{f}}_0 dt \int d^3 x \, 
\hat{\bm{j}}^\text{I}_{\text{A}_\text{P}} (t,\bm{x}) \cdot \hat{\bm{\phi}}^\text{I} (t,\bm{x}) \Bigr], 
\label{eq:UA}
\\
\hat{U}_{\text{B}_\text{Q}}
&=\text{T} \exp \Bigl[ -i \int^{t_\text{f}}_0 dt' \int d^3 y \, 
\hat{\bm{j}}^\text{I}_{\text{B}_\text{Q}} (t',\bm{y}) \cdot \hat{\bm{\phi}}^\text{I} (t',\bm{y}) \Bigr].
\label{eq:UB}
\end{align}
The local unitaries 
$\hat{U}_{\text{A}_\text{P}}$ and 
$\hat{U}_{\text{B}_\text{Q}}$
act on the Hilbert spaces 
$\mathcal{H}_\text{Ai} \otimes \mathcal{H}_{\text{F}_\text{A}}$ and 
$\mathcal{H}_\text{Bi} \otimes \mathcal{H}_{\text{F}_\text{B}}$, where 
the total Hilbert space 
$\mathcal{H}_\text{F}$ of the fields is described by 
$\mathcal{H}_\text{F} =\mathcal{H}_{\text{F}_\text{A}} \otimes \mathcal{H}_{\text{F}_\text{B}}$. 
There are no interactions induced by the fields for the factorized evolution in Eq. \eqref{eq:Udec}, which does not generate entanglement between the two objects. 

Another important point is quantum entanglement of fields' state. 
The previous work \cite{Reznik2003} showed that a pair of Unruh-DeWitt detectors, even if they are spacelike separated, becomes entangled due to the entanglement of the vacuum of a relativistic field. 
Also, there are many works about the generation of entanglement for spacelike separated detectors in the context of entanglement harvesting protocol \cite{Lin2010, Salton2015, Pozas-Kerstjens2015, Pozas-Kerstjens2016}. 
These works mean that the entanglement of the state 
$|\chi \rangle_\text{F}$ of the fields can be a source of the entanglement of the objects. 

However, in the following we find that the spacelike entanglement of fields cannot be generated in the state of the trajectories.   
The definition of entanglement as follows: a given state is not entangled if the density operator 
$\rho$ of a system has a separable form \cite{Werner1989, Nielsen2002, Horodecki2009}, 
\begin{equation}
\rho  
=  
\sum_i p_i \,  \rho_i \otimes \sigma_i,
\label{eq:rhosep}
\end{equation}
where 
$p_i$ is a probability, 
$\rho_i$ and 
$\sigma_i$ are density operators of the subsystems. 
A state which cannot be written in such form is called entangled. 
We show that the state of the objects' trajectories is written in a separable form. 
Tracing out the fields and the internal degrees of freedoms from the evolved state \eqref{eq:Psit},  for the case where the objects are in spacelike regions, the reduced density operator for the trajectories is 
\begin{equation}
\rho  
=  
\sum_{\text{P}, \text{P}' =\text{R},\text{L}} 
\sum_{\text{Q}, \text{Q}' =\text{R},\text{L}} 
\alpha_{\text{P} } \alpha^*_{\text{P}'} \beta_\text{Q}  \beta^*_{\text{Q}'} \, 
 \langle \chi' |
\hat{U}^\dagger_{\text{A}_{\text{P}'}} \hat{U}_{\text{A}_\text{P}} 
\otimes 
\hat{U}^\dagger_{\text{B}_{\text{Q}'}} \hat{U}_{\text{B}_\text{Q}} 
| \chi' \rangle \,
|\psi_\text{P} \rangle_\text{A} \langle \psi_{\text{P}'}| 
\otimes
|\psi_\text{Q} \rangle_\text{B} \langle \psi_{\text{Q}'}|,
\label{eq:rho}
\end{equation}
where we used Eq. \eqref{eq:Udec} and introduced
$|\chi' \rangle = |\chi \rangle_\text{F} |a \rangle_\text{Ai} |b \rangle_\text{Bi}$ 
as a short notation. 
The evolution operator 
$e^{-i\hat{H}_0 t_\text{f}}$ was ignored because each degree of freedom just evolves independently by the free Hamiltonian $\hat{H}_0$. 
The unitary operator
$\hat{V}^\text{A}_{\text{P}' \text{P}}=\hat{U}^\dagger_{\text{A}_{\text{P}'}} \hat{U}_{\text{A}_\text{P}}$ appearing in \eqref{eq:rho}
satisfies
\begin{equation}
\hat{V}^\text{A}_{\text{RR}}=\hat{V}^\text{A}_{\text{LL}}=\hat{\mathbb{I}}_\text{A}, 
\quad 
\hat{V}^\text{A}_{\text{LR}}=\hat{V}^{\text{A}\dagger}_{\text{RL}}=(\hat{V}^{\text{A}}_{\text{RL}})^{-1}, 
\label{eq:VA}
\end{equation}
and hence all of the unitaries 
$\hat{V}^\text{A}_{\text{RR}}, \hat{V}^\text{A}_{\text{RL}}, \hat{V}^\text{A}_{\text{LR}}$ and 
$\hat{V}^\text{A}_{\text{LL}}$ commute each other. 
This means that 
$\hat{V}^\text{A}_{\text{P}' \text{P}} $ has the following spectral decomposition, 
\begin{equation}
\hat{V}^\text{A}_{\text{P}' \text{P}} 
=\int e^{i \theta_{\text{P}'\text{P}}(\lambda)  } 
d \hat{\mu}_{\text{Ai} \otimes \text{F}_\text{A} } (\lambda) ,
\label{eq:VAspctrl}
\end{equation}
where 
$\hat{\mu}_{\text{Ai} \otimes \text{F}_\text{A} } $ 
is an operater-valued measure on the Hilbert space
$\mathcal{H}_\text{Ai} \otimes \mathcal{H}_{\text{F}_\text{A}} $. 
The real phase
$\theta_{\text{P}'\text{P}}(\lambda)$ 
has the antisymmetric property 
$\theta_{\text{P}'\text{P}}(\lambda)=-\theta_{\text{P}\text{P}'}(\lambda)$, which reflects Eq. \eqref{eq:VA}. 
As the number of trajectories for each object is two, the number of independent components of 
$\theta_{\text{P}'\text{P}} (\lambda)$ is one. 
Hence the phase is always written as 
\begin{equation}
\theta_{\text{P}'\text{P}}(\lambda)=\theta_\text{RL} (\lambda) (n_\text{P}-n_{\text{P}'}),
\label{eq:thetadec}
\end{equation}
where 
$n_\text{R}=0$ and 
$n_\text{L}=1$. 
From the above facts, we find that the reduced density operator 
$\rho$ is separable,
\begin{align}
\rho 
& 
=  
\sum_{\text{P}, \text{P}' =\text{R},\text{L}} 
\sum_{\text{Q}, \text{Q}' =\text{R},\text{L}} 
\alpha_{\text{P} } \alpha^*_{\text{P}'} \beta_\text{Q}  \beta^*_{\text{Q}'} \, 
 \langle \chi' |
\hat{V}^\text{A}_{\text{P}'\text{P}} 
\otimes 
\hat{U}^\dagger_{\text{B}_{\text{Q}'}} \hat{U}_{\text{B}_\text{Q}} 
| \chi' \rangle \,
|\psi_\text{P} \rangle_\text{A} \langle \psi_{\text{P}'}| 
\otimes
|\psi_\text{Q} \rangle_\text{B} \langle \psi_{\text{Q}'}|
\nonumber 
\\
&
=
\sum_{\text{P}, \text{P}' =\text{R},\text{L}} 
\sum_{\text{Q}, \text{Q}' =\text{R},\text{L}}
\alpha_{\text{P} } \alpha^*_{\text{P}'} \beta_\text{Q}  \beta^*_{\text{Q}'} \, 
\nonumber
\\
&
\times
\int e^{i \theta_\text{RL} (\lambda) (n_\text{P}-n_{\text{P}'})  } 
\langle \chi |
d \hat{\mu}_{\text{Ai} \otimes \text{F}_\text{A}} (\lambda) \otimes \hat{U}^\dagger_{\text{B}_{\text{Q}'}} \hat{U}_{\text{B}_\text{Q}}  | \chi \rangle_\text{F} \,
|\psi_\text{P} \rangle_\text{A} \langle \psi_{\text{P}'}| 
\otimes
|\psi_\text{Q} \rangle_\text{B} \langle \psi_{\text{Q}'}| 
\nonumber 
\\
&
=
\int d\mu(\lambda) \,
|\psi(\lambda) \rangle_\text{A} \langle \psi (\lambda) | \otimes \sigma_\text{B} (\lambda),
\label{eq:rhosep}
\end{align}
where we used Eqs. \eqref{eq:VAspctrl} and \eqref{eq:thetadec} and defined the probability measure 
$\mu$ with
$d\mu(\lambda)= \langle \chi' | d\hat{\mu}_{\text{Ai} \otimes \text{F}_\text{A}}(\lambda) |\chi' \rangle$, the state
$|\psi (\lambda) \rangle_\text{A}$ and the density operator
$\sigma_\text{B} (\lambda)$ as 
\begin{align}
|\psi (\lambda) \rangle_\text{A}
&=\sum_{\text{P} =\text{R},\text{L}} \alpha_\text{P} e^{i \theta_\text{RL} (\lambda) n_\text{P}} |\psi_\text{P} \rangle_\text{A}, \label{eq:psi}
\\
\sigma_\text{B} (\lambda)
&=\frac{1}{d\mu (\lambda)} \sum_{\text{Q}, \text{Q}' =\text{R},\text{L}}
\beta_\text{Q}  \beta^*_{\text{Q}'} \langle \chi' |
d \hat{\mu}_{\text{Ai} \otimes \text{F}_\text{A}} (\lambda) \otimes \hat{U}^\dagger_{\text{B}_{\text{Q}'}} \hat{U}_{\text{B}_\text{Q}}  | \chi' \rangle \,
|\psi_\text{Q} \rangle_\text{B} \langle \psi_{\text{Q}'}|.
\label{eq:phi}
\end{align}
Here, we emphasize that the Hilbert space 
$\mathcal{H}_\text{F}$ of the fields has no negative norm states, which was mentioned below Eq. \eqref{eq:Htot}. 
The fact leads to the inequalities
$\mu(\lambda) \geq 0$ 
and 
$\sigma_\text{B} (\lambda) \geq 0$ and gurantees that 
$\mu (\lambda) $ 
and 
$\sigma_\text{B} (\lambda)$ are a probability measure and a density operator, respectively. 
Hence the separablity of the state of the objects' trajectories holds. 
If gauge degrees of freedom are included in the fields, the Hilbert space 
$\mathcal{H}_\text{F}$ may have a negative norm state and the separability is not always guaranteed.

The separability of the objects does not depend on the dynamics of fields and the details of classical trajectories. 
Also, the seprability holds even for the case where the objects' state of the internal degrees of freedom and the fields are initially in a mixed state. 
Our result means that the fields do not play a role of quantum mediators to generate the spacelike entanglement among the trajectories of such objects. 

We compare our result with the no-go theorems in \cite{Marletto2017, Simidzija2018} on generation of entanglement. 
The theorem in \cite{Marletto2017} argued that two systems mediated by classical systems with only a single observable (this is the meaning of ``classical'' for that claim) have no entanglement. 
For our model, the mediators are the fields, which may have noncommutative observables, for example, the field operator and its conjugate. 
In this sense, the fields can be quantum systems in general. 
However, there are no generations of spacelike entanglement. 

The no-go theorem in Ref. \cite{Simidzija2018} elucidates our result. 
We can rewrite Eq. \eqref{eq:Psit} for the spacelike separated two objects as 
\begin{align}
| \Psi_\text{f} \rangle 
&
=
e^{-i \hat{H}_0 t_\text{f} }  
\sum_{\text{P}, \text{Q}=\text{R},\text{L}} 
\alpha_\text{P} \, \beta_\text{Q} |\psi_\text{P} \rangle_\text{A} | \psi_\text{Q} \rangle_\text{B} 
\otimes \hat{U}_\text{PQ}| \chi \rangle_\text{F} |a \rangle_\text{Ai} |b \rangle_\text{Bi}
\nonumber 
\\
&
=
e^{-i \hat{H}_0 t_\text{f} }  
\sum_{\text{P}, \text{Q}=\text{R},\text{L}} 
\alpha_\text{P} \, \beta_\text{Q} |\psi_\text{P} \rangle_\text{A} | \psi_\text{Q} \rangle_\text{B} 
\otimes (\hat{U}_{\text{A}_\text{P}} \otimes \hat{U}_{\text{B}_\text{Q}} ) | \chi \rangle_\text{F} |a \rangle_\text{Ai} |b \rangle_\text{Bi}
\nonumber 
\\
&
=
e^{-i \hat{H}_0 t_\text{f} }  
\Bigl(
 \sum_{\text{P}=\text{R},\text{L}} | \psi_\text{P} \rangle_\text{A} \langle \psi_\text{P} |
  \otimes 
  \hat{U}_{\text{A}_\text{P}} 
  \otimes 
  \hat{\mathbb{I}}_{\text{F}_\text{B}}
 \otimes 
  \hat{\mathbb{I}}_\text{B} 
\Bigr)
\otimes
\Bigl(
 \hat{\mathbb{I}}_\text{A} 
 \otimes
 \sum_{\text{Q}=\text{R},\text{L}} | \psi_\text{Q} \rangle_\text{B} \langle \psi_\text{Q} | 
 \otimes 
 \hat{\mathbb{I}}_{\text{F}_\text{A}}
 \otimes 
 \hat{U}_{\text{B}_\text{Q}}
\Bigr)
|\Psi_\text{in} \rangle,
\label{eq:Psit3}
\end{align}
where we used Eq. \eqref{eq:Udec}, and 
$|\Psi_\text{in} \rangle$ is the initial state given in Eq. \eqref{eq:Psiin}. 
In this formula, we find the controlled unitary
$\hat{U}_\text{AF}$, 
\begin{equation}
\hat{U}_\text{AF}= 
\sum_{\text{P}=\text{R},\text{L}} |\psi_\text{P} \rangle_\text{A} \langle \psi_\text{P} | 
\otimes \hat{U}_{\text{A}_\text{P}} 
\otimes \hat{\mathbb{I}}_{\text{F}_\text{B}}. 
\label{eq:cU}
\end{equation}
Exactly speaking, 
$\hat{U}_\text{AF}$ has inverse only when it acts on the subspace spanned by 
$|\psi_\text{R} \rangle_\text{A}$ and 
$|\psi_\text{L} \rangle_\text{A}$ of the Hilbert space 
$\mathcal{H}_\text{A}$. 
In Ref. \cite{Simidzija2018}, the authors showed that the unitary evolution 
$\hat{U}= (\hat{U}_\text{AS}\otimes \hat{\mathbb{I}}_\text{B})(\hat{\mathbb{I}}_\text{A} \otimes \hat{U}_\text{BS} ) $ with
the exponential of a Schmidt rank-1 operator 
$\hat{U}_\text{AS}=e^{-i\hat{m}_\text{A} \otimes \hat{X}_\text{S}}$ 
does not generate entanglement between the systems A and B. 
The systems A, B and S correspond to the objects A and B, and the fields F for our model.
The controlled unitary
$\hat{U}_\text{AF}$ is rewritten as the form 
\begin{align}
\hat{U}_\text{AF}=\hat{U}_{\text{A}_\text{R}} (|\psi_\text{R} \rangle_\text{A} \langle \psi_\text{R} | 
\otimes \hat{\mathbb{I}}_{\text{F}_\text{A}}
\otimes \hat{\mathbb{I}}_{\text{F}_\text{B}}
+
|\psi_\text{L} \rangle_\text{A} \langle \psi_\text{L} | 
\otimes \hat{V}^\text{A}_\text{RL}
\otimes \hat{\mathbb{I}}_{\text{F}_\text{B}})
=\hat{U}_{\text{A}_\text{R}} e^{-i\hat{m}_\text{A} \otimes \hat{X}_\text{F}},
\end{align}
where 
$\hat{V}^\text{A}_\text{RL} = \hat{U}^\dagger_{\text{A}_\text{R}} \hat{U}_{\text{A}_\text{L}}$, and the self-adjoint operator 
$\hat{X}_\text{F}$ satisfies 
$e^{-i \hat{X}_\text{F}}=\hat{V}^\text{A}_\text{RL}
\otimes \hat{\mathbb{I}}_{\text{F}_\text{B}}$, and 
$\hat{m}_\text{A}= 0 \times  |\psi_\text{R} \rangle_\text{A} \langle \psi_\text{R} | 
+1 \times |\psi_\text{L} \rangle_\text{A} \langle \psi_\text{L} | $.
Since the entanglement between the two objects is invariant under the local unitary transformation $\hat{U}_{\text{A}_\text{R}}$, 
the controlled unitary 
$\hat{U}_\text{AF}$ plays the same role as the exponetial of a Schmidt rank-1 operator. 
Thus, our no-go result on generation of spacelike entanglement is a consequence of the no-go theorem in \cite{Simidzija2018}. 
Note that the no-go theorem can be applied under the approximation assigning local currents \eqref{eq:jAjB} and for the states of trajectories satisfying 
$\langle \psi_\text{P} | \psi_{\text{P}'} \rangle \approx \delta_{\text{PP}'}$. 
If these conditions do not hold, we need a further study on entanglement generation.

We comment on the extension of our model.
It is well known that the spacelike entanglement of a field is extracted by the Unruh-DeWitt detectors \cite{Reznik2003}. 
Further, in Refs. \cite{Foo2021, Henderson2020}, the authors discussed an entanglement harvesting protocol by using the Unruh-DeWitt detectors with quantum superpositions of trajectories. 
The critical difference is that the states of trajectories are only focused on to show the separability. 
This means that the information of internal degrees of freedom are neccessary for an extraction of spacelike entanglement from fields.
Further, it is worth considering a multi-partite \cite{Miki2020} or multi-trajectory \cite{Tilly2021} extended model of the QGEM proposal, since our result is based on the fact that each of two objects is superposed in two classical trajectories. 
It is interesting to characterize the advantage of many objects or trajectories for the generation of spacelike entanglement of fields.  

\section{Conclusion}
\label{sec:Conclusion}

In the QGEM proposal, it was demonstrated that two spatially superposed objects can be a probe of gravity-induced entanglement. 
We discussed how such objects probe state entanglement of quantum fields.
We considered a pair of objects in a superposition of local states which couple with quantum fields. 
In this system, there are no constraints for the whole system and the fields have only dynamical degrees of freedom.  
From the entanglement analysis for the objects with the approximated currents evaluated on each trajectory, we found that the state of the trajectories cannot be entangled if the objects are in spacelike regions. 
This result is independent of the dynamics of fields and the detail of objects' trajectories, which holds if the commutator of fields vanishes for spacelike separated regions (microcausality). 
The limitation for entanglement generation characterizes the behavior of the fields as quantum mediators between the two superposed objects. 
In other words, the position space of such objects cannot store the spacelike entanglement of fields.  
We can imagine several strategies ; use of the information of trajectories and internal degrees of freedom, and extensions with multiple objects and object superposed in multiple trajectories.
It is important to discuss how the extensions are effective for the detection of spacelike entanglement.  
We need further research on quantum objects which play a crucial role in probing quantum nature of fields.

\begin{acknowledgments}
We thank A. Mazumdar, J. Soda, S. Kanno, and K. Yamamoto for useful discussions and comments related to this paper. 
\end{acknowledgments}
\begin{appendix}
\end{appendix}



\begin{thebibliography}{10}
\newcommand{\enquote}[1]{``#1''}

\bibitem{Feynmann1995}
R. P. Feynmann, F. M. Morinigo and W. G. Wagner, \emph{Feynmann lectures on gravitation},
(Westview Press, 1995)

\bibitem{Aharony2000}
O. Aharony, S. S. Gubser, J. Maldacena, H. Ooguri and Y. Oz, \enquote{Large N Field Theories, String
Theory and Gravity}, \emph{Phys. Rept.} \textbf{323}, 183 (2000)

\bibitem{Kiefer2006}
C. Kiefer, \enquote{Quantum gravity: general introduction and recent developments}, \emph{Ann. Phys.} \textbf{15}, 129 (2006)

\bibitem{Woodard2009}
R. P. Woodard, \enquote{How far are we from the quantum theory of gravity?}, \emph{Rep. Prog. Phys.} \textbf{72}, 126002 (2009)

\bibitem{Doherty2013}
M. W. Doherty, N. B. Manson, P. Delaney, F. Jelezko, J. Wrachtrup and L. C. L. Hollenberg, \enquote{The nitrogen-vacancy colour centre in diamond}, \emph{Phys. Rept. } \textbf{528}, 1 (2013).

\bibitem{Aspelmeyer2014}
M. Aspelmeyer, T. J. Kippenberg  and F. Marquardt, \enquote{Cavity optomechanics} \emph{Rev. Mod. Phys.} \textbf{86} 1391–452 (2014). 

\bibitem{Matsumoto2019} 
N. Matsumoto, S. B. Cata$\tilde{\text{n}}$o-Lopez, M. Sugawara, S. Suzuki, N. Abe, K. Komori, Y.
Michimura, Y. Aso, and K. Edamatsu, \enquote{Demonstration of Displacement Sensing of a mgScale Pendulum for mm- and mg-Scale Gravity Measurements}, \emph{Phys. Rev. Lett.} \textbf{122}, 071101 (2019).

\bibitem{Catano-Lopez2020}
 S. B. Cata$\tilde{\text{n}}$o-Lopez, J. G. Santiago-Condori, K. Edamatsu, and N. Matsumoto, 
 \enquote{High-Q Milligram-Scale Monolithic Pendulum for Quantum-Limited Gravity Measurements}, \emph{Phys.
Rev. Lett.} \textbf{124}, 221102 (2020).


\bibitem{Carney2018}
 D. Carney, P. C. E. Stamp and J. M. Taylor, \enquote{Tabletop experiments for quantum gravity: a user’s manual}, \emph{Class. Quant. Grav.} \textbf{36}, 034001 (2018).  
 
\bibitem{Bose2017}
S. Bose, A. Mazumdar, G. W. Morley, H. Ulbricht, M Toro$\check{\text{s}}$, M. Paternostro, A. A. Geraci, P. F. Barker, M. S. Kim, and G. Milburn, \enquote{Spin Entanglement Witness for Quantum Gravity}, \emph{Phys. Rev. Lett.} \textbf{119},  240401 (2017).

\bibitem{Marletto2017}
C. Marletto and V. Vedral, \enquote{Gravitationally Induced Entanglement between Two Massive Particles is Sufficient Evidence of Quantum Effects in Gravity}, \emph{Phys. Rev. Lett.} \textbf{119},  240402 (2017).

\bibitem{Marletto2018}
C. Marletto and V. Vedral, \enquote{When can gravity path-entangle two spatially superposed masses?}, \emph{Phys. Rev. D} \textbf{98},  046001 (2018).

\bibitem{Carlesso2019}
M. Carlesso, A. Bassi, M. Paternostro, and H. Ulbricht, \enquote{Testing the gravitational field generated by a quantum superposition}, \emph{New J. Phys.} \textbf{21}, 093052 (2019).

\bibitem{Krisnanda2020}
T. Krisnanda, G. Y. Tham, M. Paternostro, and T. Paterek, \enquote{Observable quantum entanglement due to gravity} \emph{npj Quant. Inf.} \textbf{6}, 12 (2020).

\bibitem{vandeKamp2020}
T. W. van de Kamp, R. J. Marshman, S. Bose, and A. Mazumdar, \enquote{Quantum gravity witness via entanglement of masses: Casimir screening}, \emph{Phys. Rev. A} \textbf{102}, 062807 (2020).

\bibitem{Grobardt2020} 
A. Großardt, \enquote{Acceleration noise constraints on gravity-induced entanglement}, \emph{Phys. Rev. A} \textbf{102}, 040202(R) (2020).

\bibitem{Qvafort2020}
S. Qvarfort, S. Bose, and A. Serafini, \enquote{Mesoscopic entanglement through central–potential interactions}, \emph{J. Phys. B} \textbf{53}, 235501 (2020).

\bibitem{Guerreiro2020}
T. Guerreiro, \enquote{Quantum Effects in Gravity Waves}, \emph{Class. Quant. Grav.} \textbf{37}, 155001 (2020)

\bibitem{Kanno2020}
S. Kanno, J. Soda, and J. Tokuda, \enquote{Noise and decoherence induced by gravitons}, 
\emph{Phys. Rev. D} \textbf{103}, 044017 (2021).

\bibitem{Hall2018}
M. J. Hall and M. Reginatto, \enquote{On two recent proposals for witnessing nonclassical gravity}, \emph{J. Phys. A} \textbf{51}, 085303 (2018).

\bibitem{Belenchia2018}
A. Belenchia, R. M. Wald, F. Giacomini, E. Castro-Ruiz, $\check{\text{C}}$. Brukner, and M. Aspelmeyer, \enquote{Quantum superposition of massive objects and the quantization of gravity}, \emph{Phys. Rev.D} \textbf{98}, 126009 (2018).

\bibitem{Marshman2020}
R. J. Marshman, A. Mazumdar, and S.Bose, \enquote{Locality and entanglement in table-top testing of the quantum nature of linearized gravity}, \emph{Phys. Rev. A} \textbf{101}, 052110 (2020).

\bibitem{Balushi2018}
A. A. Balushi, W. Cong, and R. B. Mann, \enquote{Optomechanical quantum Cavendish experiment}, \emph{Phys. Rev. A} \textbf{98}, (2018) 043811.

\bibitem{Miao2020}
H. Miao, D. Martynov, H. Yang, and A. Datta, \enquote{Quantum correlations of light mediated by gravity}, \emph{Phys. Rev. A} \textbf{101}, (2020) 063804.

\bibitem{Matsumura2020}
A. Matsumura and K. Yamamoto, \enquote{Gravity-induced entanglement in optomechanical systems}, \emph{Phys. Rev. D} \textbf{102}, 106021 (2020).

\bibitem{Howl2019}
R. Howl, R. Penrose and I, Fuentes, \enquote{Exploring the unification of quantum theory and general relativity with a Bose–Einstein condensate}, \emph{New J. Phys.} \textbf{21} 043047 (2019). 

\bibitem{Anastopoulos2020}
C. Anastopoulos and B.-L. Hu, \enquote{Quantum superposition of two gravitational cat states}, \emph{Class. Quant. Grav.} \textbf{37}, 235012, (2020).

\bibitem{Nguyen2020}
H. C. Nguyen and F. Bernards, \enquote{Entanglement dynamics of two mesoscopic objects with gravitational interaction}, \emph{Eur. Phys. J. D} \textbf{74}, 69 (2020).

\bibitem{Miki2020}
D. Miki, A. Matsumura and K. Yamamoto, \enquote{Entanglement and decoherence of massive particles due to gravity},  	arXiv:2010.05159  (2020).

\bibitem{Tilly2021}
J. Tilly, R. J. Marshman, A. Mazumdar and S. Bose, \enquote{Qudits for Witnessing Quantum Gravity Induced Entanglement of Masses Under Decoherence}, arxiv:2101.08086.

\bibitem{Reznik2003}
B. Reznik, \enquote{Entanglement from the vacuum}, \emph{Found. Phys.} \textbf{33}, 167 (2003).

\bibitem{Lin2010} 
S.-Y. Lin and B. L. Hu, \enquote{Entanglement creation between two causally disconnected objects},  \emph{Phys. Rev. D} \textbf{81}, 045019 (2010).


\bibitem{Salton2015}
G. Salton, R. B. Mann, and N. C. Menicucci, \enquote{Acceleration-assisted entanglement harvesting and rangefinding}, \emph{New J. Phys.} \textbf{17}, 035001 (2015).

\bibitem{Pozas-Kerstjens2015} 
A. Pozas-Kerstjens and E. Mart\'{i}n-Mart\'{i}nez, \enquote{Harvesting correlations from the quantum vacuum},  \emph{Phys. Rev. D} \textbf{92}, 064042 (2015).

\bibitem{Pozas-Kerstjens2016} 
A. Pozas-Kerstjens and E. Mart\'{i}n-Mart\'{i}nez, \enquote{Entanglement harvesting from the electromagnetic vacuum with hydrogenlike atoms},  \emph{Phys. Rev. D} \textbf{94}, 064074 (2016).

\bibitem{Simidzija2018a} 
P. Simidzija, and E. Mart\'{i}n-Mart\'{i}nez, \enquote{Harvesting correlations from thermal and squeezed coherent states},  \emph{Phys. Rev. D} \textbf{98}, 085007 (2018).

\bibitem{Stritzelberger2021} 
N. Stritzelberger, L. J. Henderson, V. Baccetti, N. C. Menicucci, and A. Kempf \enquote{Entanglement harvesting with coherently delocalized matter},  \emph{Phys. Rev. D} \textbf{103}, 016007 (2021).

\bibitem{Henderson2020} 
L. J. Henderson, A. Belenchia, E. Castro-Ruiz, C. Budroni, M. Zych, $\check{\text{C}}$. Brukner, and R. B. Mann, \enquote{Quantum Temporal Superposition: The Case of Quantum Field Theory},  \emph{Phys. Rev. Lett.} \textbf{125}, 131602 (2020).

\bibitem{Foo2021} 
J. Foo, R. B. Mann, M. Zych, \enquote{Entanglement amplification between superposed detectors in flat and curved spacetimes},  arXiv:2101.01912 (2021).

\bibitem{Ford1993}
L. H. Ford, \enquote{Electromagnetic vacuum fluctuations and electron coherence}, \emph{Phys. Rev. D} \textbf{47},  5571 (1993).

\bibitem{Breuer2001}
H.-P. Breuer and F. Petruccione, \enquote{Destruction of quantum coherence through emission of bremsstrahlung}, \emph{Phys. Rev. A} \textbf{63}, 032102 (2001).

\bibitem{Werner1989}
R. Werner, \enquote{Quantum states with Einstein-Podolsky-Rosen correlations admitting a hidden-variable model}, \emph{Phys. Rev. A} \textbf{40}, (1989) 4277.
  
\bibitem{Nielsen2002}
M. A. Nielsen and I. Chuang, \emph{Quantum Computation and Quantum Information} (Cambridge University Press, Cambridge, England, 2002).

\bibitem{Horodecki2009}
R. Horodecki, P. Horodecki, M. Horodecki, and K. Horodecki, \enquote{Quantum entanglement}, \emph{Rev. Mod. Phys.} \textbf{81}, 865 (2009).

\bibitem{Weinberg1995}
See, e.g. S. Weinberg, \emph{The Quantum Theory of Fields, Volume I: Foundations} (Cambridge University Press, Cambridge,
England, 1995)

\bibitem{Weinberg1996}
See, e.g., S. Weinberg, \emph{The Quantum Theory of Fields, Volume II, Modern Applications} (University Press,
Cambridge, UK, 1996)

\bibitem{Simidzija2018} 
P. Simidzija, R. H. Jonsson, and E. Mart\'{i}n-Mart\'{i}nez, \enquote{General no-go theorem for entanglement extraction},  \emph{Phys. Rev. D} \textbf{97}, 125002 (2018).

















\end{thebibliography}
\end{document}